# Functionalization of Single-Wall Carbon Nanotubes with Chromophores of Opposite Internal Dipole Orientation

Yuanchun Zhao,[†] Changshui Huang,[‡] Myungwoong Kim,[‡] Bryan M. Wong, [⊥, §]

François Léonard,[⊥] Padma Gopalan,[*,‡] and Mark A. Eriksson[*,†]

[†]Department of Physics, University of Wisconsin-Madison, Madison, Wisconsin 53706, USA

[‡]Department of Materials Science and Engineering, University of Wisconsin-Madison, Madison, Wisconsin 53706, USA

[⊥]Sandia National Laboratories, Livermore, California 94551, USA

**ABSTRACT:** We report the functionalization of carbon nanotubes with two azobenzene-based chromophores with large internal dipole moments and opposite dipole orientations. The molecules are attached to the nanotubes non-covalently via a pyrene tether. A combination of characterization techniques shows uniform molecular coverage on the nanotubes, with minimal aggregation of excess chromophores on the substrate. The large on/off ratios and the sub-threshold swings of the nanotube-based field-effect transistors (FETs) are preserved after functionalization, and different shifts in threshold voltage are observed for each chromophore. *Ab initio* calculations verify the properties of the synthesized chromophores and indicate very small charge transfer, confirming a strong, non-covalent functionalization.

1. INTRODUCTION

Directing and assembling dipolar molecules on surfaces is of fundamental significance for nano/micro-scale systems,[1-3] offering opportunities for optoelectronic devices[4-7] and spintronics.[8] Single-wall carbon nanotubes (SWNTs) consisting of a single layer of carbon atoms are highly sensitive to the surrounding electrostatic environment, making them ideal probes of nearby molecules.[9-11] The electronic and optical properties of SWNTs make them promising building blocks for future nanoelectronic and optoelectronic devices,[12] and functionalization is a valuable strategy to tune and optimize their properties.[13-16] As a result, many research groups have studied the use of spiropyrans,[9,17] porphyrins,[18,19] and dipolar molecules based on azobenzenes,[5,20-24] to functionalize SWNTs.

Previous work has demonstrated optically active SWNT field-effect transistors (FETs) by noncovalently functionalizing the nanotube with azobenzene-based chromophores.[5,20] In that work the optical activity was attributed to the photoisomerization and concurrent dipole moment change in the chromophore molecules. Such effects should depend on the chromophore dipole orientation, and previous measurements of optical second harmonic generation (SHG) are consistent with a net molecular orientation and a non-zero average tilt angle of the choromphores away from the primary nanotube axis of symmetry and towards the normal of the underlying substrate.[23] An ability to cleanly functionalize nanotubes with tailored chromophores, e.g., to control the direction and magnitude of the dipole moments, would provide an important tool for device design, engineering, and optimization. For example, the ability to shift threshold voltages to either more negative or more positive directions would provide significant design flexibility, offering the potential to engineer electro-optic detectors with zero current flow (and thus reduced power consumption) in both the 'off' and the 'on' states, in analogy with CMOS electronics.

Here we report the fabrication and characterization of SWNT transistor-like devices that are non-covalently functionalized with two closely related pyrene-tethered molecules with large internal dipole moments: i) a Disperse Red 1 (DR1) derivative (DR1P) that has been used for this purpose in the past[5,20,23,24], and ii) a newly designed and synthesized azobenzene chromophore (DR1P') having a dipole moment in the opposite direction from DR1P when anchored to the SWNT. *Ab initio* calculations were used to obtain the structure and dipole moments of these molecules, and the results show that the dipole moments are oriented with the positive center on opposite ends of DR1P and DR1P'. Raman spectroscopy of the nanotubes on the substrate indicates that both DR1P and DR1P' can be successfully bound to the nanotubes with little disruption of the nanotube properties. We confirm with atomic force microscopy (AFM) that the functionalization is uniform along the length of nanotubes and that the substrate itself does not collect aggregates of chromophores. *Ab initio* calculations of the structure of the chromophores on the nanotubes are consistent with these experimental observations. Comparisons of field-effect transistors before and after functionalization show no degradation of on current, on/off ratio, or subthreshold swing, and indicate that DR1P and DR1P' affect the threshold voltage ($V_{th}$) of the SWNT devices differently.

## 2. EXPERIMENTAL SECTION

**2.1. Design and synthesis of chromophores**. We used DR1, a well-studied and commercially available azobenzene chromophore, as a starting molecule to build dipolar functionalities on SWNTs. This molecule was directly coupled with pyrene butyric acid, resulting in DR1P, as we have reported previously.[5,20,24] To achieve the opposite dipole moment orientation, we designed a new azobenzene derivative, which we label as DR1P' (compound 4 in Scheme 1), following the procedure in scheme 1. Compound 1 was synthesized via a straightforward diazonium coupling reaction.[22,25] The resulting product was reacted with compound 2 (synthesized by an acid-catalyzed Fischer esterification of cyanoacetic acid and ethylene glycol) via the Knoevenagel condensation reaction,[26] resulting in DROH. The terminal hydroxyl group in DROH was further coupled with pyrene butyric acid through Steglich esterification,[5,20] leading to DR1P'. A UV/vis absorption spectrum of 5 µM DR1P' solution in chloroform shows the expected peaks from pyrene unit in the 300 ~ 400 nm range and a strong absorption in the visible with $\lambda_{max}$ ~ 525 nm, corresponding to the absorption peak of π-conjugated azobenzene derivative with donor and acceptor groups.[26] (see Supporting Information for additional details of the synthesis as well as NMR and UV/vis data.)

2.2 **Computational Details.** All density functional theory (DFT) calculations were performed using the M06-L functional, which is designed specifically for noncovalent and $\pi$-$\pi$ stacking interactions.[27] While it is well-known that the choice of DFT functional strongly affects both the interaction energy and charge-transfer properties in functionalized SWNTs,[28] we specifically chose the M06-L functional since it is a semi-local DFT method that can be used routinely for the large periodic systems (over 200 atoms) studied here. Geometry optimizations were carried out for all systems, and a large 6-31G(d,p) basis set was used to calculate noncovalent interaction energies between the chromophores and the SWNTs. In order to provide further insight into the electronic properties of these functionalized SWNTs, we also computed the density of states at the same M06-L/6-31G(d,p) level of theory (see Supporting Information, Figure S5). For all of the DFT calculations, a (10,0) semiconducting SWNT (diameter = 7.9 Angstroms) was chosen as a representative model system, and calculations were performed using a one-dimensional supercell along the axis of the nanotube. Since the chromophore molecules are over 6 times longer than the (10,0) SWNT unit cell, a large supercell of 17.1 Å along the nanotube axis was chosen to allow adequate separation between adjacent chromophores. Both the geometry and the unit cell of the chromophore-functionalized SWNTs were optimized without constraints, resulting in a $\pi$-$\pi$ stacking distance of 3.3 Angstroms between the chromophore and the SWNT. All calculations were carried out with a locally modified version of the Gaussian software code.

2.3 **Fabrication of SWNT field-effect transistors (FETs).** The nanotubes used in the experimental work were purchased from NanoIntegris Inc. (IsoNanotube-S 95%) and were received in powder form with 95%-enriched semiconducting nanotubes. Nanotube powders of 0.1 mg were dispersed and sonicated in 10 ml o-dichlorobenzene (ODCB) using a bath sonicator for 2 h, followed by centrifugation (11,000 rpm, 3 h) to remove large bundles. The supernatant solution was decanted and spun coated on Si substrates with a 200 nm thermal oxide layer. The density of nanotubes was controlled by dilution of the solution. The source/drain electrodes are formed from a three-layer structure of Ti/Pd/Au (0.5 nm/15 nm/25 nm)[29] and fabricated by electron beam lithography (EBL), metal evaporation, and liftoff. The nanotube channel length ($L_{ch}$) is 500 nm.

**2.4 Functionalization.** Both DR1P and DR1P' were used to functionalize SWNTs. For functionalization, the device chips were soaked in 0.5 mM chromophore (DR1P or DR1P') solutions in methylene chloride for 24 hours. The chromophore molecules were noncovalently attached on the nanotubes through the pyrene tether.[30] After soaking, they were washed with methylene chloride and methanol to remove any physisorbed molecules.[24] As a control pyrene (the tether part for DR1P and DR1P') was also used to functionalize SWNTs using an identical process.

**2.5 Characterization and measurements.** The chromophores and the chromophore-nanotube hybrids deposited on oxidized Si substrates were characterized by Raman Spectroscopy (Horiba Jobin Yvon LabRAM ARAMIS) using a 532 nm laser excitation. SWNTs spun-coated on oxidized Si substrates were measured by atomic force microscopy (AFM, Nanoscope IIIa) before and after functionalization. Cross-sectional analysis of the images enables the extraction of the nanotube diameter and the change in observed height arising from the adsorption of the chromophores studied. Current-gate voltage ($I$-$V_g$) characteristics of the individual nanotube

FETs were measured with a source-drain bias of 100 mV, and the heavily doped Si substrate was used as a backgate.

## 3. RESULTS AND DISCUSSION

Figure 1a and 1b show the optimized geometries from DFT calculations of the newly designed DR1P' attached on a (10,0) nanotube with perpendicular and parallel orientations, respectively. To verify the notion of dipole reversal of the designed chromophores, we performed *ab initio* calculations on the isolated molecules by optimizing the molecular geometries at the M06-L/6-31G(d,p) level of theory (see Computational Details section). From these optimized geometries, we calculated the electric dipole moments of the isolated chromophores as expectation values of the dipole operator using the Kohn-Sham density matrix. As shown in Figure 1c and 1d, the dipole moment of each chromophore is found to lie parallel to the azobenzene functional group. DR1P and DR1P' were found to have opposite dipole moment orientations with respect to the coupled pyrene tether. The calculations show that DR1P has a dipole moment of 14.1 D, defined here to be positive, and DR1P' has a dipole moment in the opposite sense of -10.0 D.

Figures 2a and 2b show AFM images of an individual SWNT before and after functionalization with DR1P. After functionalization, both the nanotube and the substrate surface exhibit very similar morphologies to those of the pristine sample, and no aggregation of chromophores was observed. The only difference observed is the increase in the apparent nanotube diameter by ~ 0.3 nm after functionalization (Figure 2c and 2d), which corresponds to a typical π-π stacking distance.[31] The standard deviation of the AFM height measurements at different positions on the SWNT is 0.1 nm and remains essentially unchanged after functionalization, indicating that the coverage on the nanotube is uniform. To further confirm functionalization with the chromophore, we performed Raman spectroscopy on SWNT network samples using a 532 nm laser excitation. As a reference, the inset of Figure 2e shows the Raman spectra of the chromophores alone (no nanotubes) deposited by drop casting onto an oxidized Si substrate. Because of their similar molecular structures, DR1P and DR1P' show similar Raman spectra, except for peaks near 1088 cm$^{-1}$ and 1330 cm$^{-1}$ that are present in the DR1P spectrum but absent from that of DR1P'. The former peak is attributed to C–N$_{(NO)}$ stretching vibration and the latter is related to the N–O stretching mode.[32] Figure 2e shows the Raman spectra of the pristine SWNTs and those functionalized with DR1P and DR1P'. Weak but distinct Raman signals of DR1P and DR1P' were detected in the functionalized SWNTs. Because of the 95% semiconducting enriched nanotubes used in this study, the Raman G band of the pristine SWNTs exhibits a narrow lineshape.[33] After functionalization, both the lineshape and frequency of the G band remained unchanged, indicating the lack of significant charge transfer between the chromophores and SWNTs,[34,35] consistent with non-covalent functionalization.

The above results raise interesting questions regarding the structure of the chromophores on the nanotubes. For example, the AFM results indicate a height increase of 0.3 nm, which is much smaller than the linker length or azobenzene units of the chromophores. To address this question, we carried out large-scale (periodic) DFT calculations for each of the composite chromophore-SWNT structures at the same M06-L/6-31G(d,p) level of theory as for the isolated chromophores, where a (10,0) semiconducting SWNT was chosen as a representative model system. We calculated several geometrical orientations of both DR1P and DR1P' on the nanotube, two of which are shown in Figure 1a and 1b. Among all of the various configurations, we find that the most favorable binding interaction occurs when the chromophores lie parallel to the SWNT axis (i.e., when both the pyrene linker and the chromophore have π-stacking interactions with the SWNT, Figure 1b). The noncovalent interaction energy for DR1P in the parallel configuration is 1.55 eV (we use the sign convention that a positive interaction energy indicates an attractive/favorable interaction), which is slightly smaller than the 1.74 eV noncovalent interaction energy for DR1P' (DR1P' possesses more functional groups that interact with the SWNT compared to the smaller DR1P chromophore). As a comparison, the noncovalent interaction energies for the perpendicular orientations were found to be about 0.91 eV smaller. These results indicate that significant molecular distortions can arise in these systems despite the relatively short pyrene-azobenzene linker. The stronger interaction energy of the parallel configurations suggests a small, π-stacking increase of the nanotube size upon functionalization, in agreement with the AFM measurements.

We now turn to measurements of current through nanotube-based FET devices (Figure 1e), making direct comparisons of the same nanotube devices before and after functionalization. Figure 3a-c show measurements of the current $I_{ds}$ as a function of the gate voltage $V_g$ through individual nanotube FETs both before and after functionalization with DR1P, DR1P', and (as a control) pyrene, respectively. The data presented were acquired after continuously sweeping $V_g$ until the curves overlap repeatedly. This procedure enables results repeatable enough that reliable comparison can be made of the same nanotube FETs in ambient conditions both before and after functionalization. All the FETs we discuss here show *p*-type transport behavior. The relatively short channel length $L_{ch} = 500$ nm leads to a large on-state current (typically $> 0.5$ μA) for a source-drain bias $V_{ds} = 100$ mV.

As shown in Figure 3a-c, we observe on/off ratios that exceed $10^4$, and more importantly, the on/off ratios are preserved after functionalization for each of the three molecules studied here. In addition, we find little or no degradation of the on current. Figure 3d summarizes the transistor sub-threshold swings, the voltage required to change the current by one order of magnitude, for the nanotube FETs, both before and after functionalization. In total, we have measured 23 nanotube devices on six different chips, two chips for each type of functionalization. In all cases the functionalization has the effect of making the sub-threshold swing more consistent on the forward and backward gate voltage sweeps, and the ratio of the standard deviation to the mean of the sub-threshold swings is unity before functionalization and 0.4 after, indicating an improvement in device uniformity.

From plots of the same data as in Figure 3a-c but on linear ordinate axes (not shown), we can extract the shift in the threshold voltage $V_{th}$ after functionalization for each device. These data are summarized for both the backward and forward gate voltage sweeps in Figure 3e and 3f. In comparison to the pristine SWNTs, the DR1P functionalized devices show negative threshold shift $\Delta V_{th}$ in both the backward and forward $V_g$ sweeps of -12 V and -2.6 V, respectively. In contrast, the backward sweeps for DR1P' exhibit a negative $\Delta V_{th}$, the same direction as for DR1P, but of more than 3 times smaller magnitude. The forward $V_g$ sweeps after functionalization with DR1P' show essentially no $\Delta V_{th}$. In order to test what fraction of these effects arises from the chromophore itself, we compared with measurements of pyrene-functionalized nanotube devices, for which a negative $\Delta V_{th}$ is observed in both the forward and backward $V_g$ sweeps. The shifts for the pyrene functionalized devices are smaller than those of the DR1P devices, but they are slightly more pronounced than those of the DR1P' devices.

To discuss the magnitude of these threshold voltage shifts, we first estimate the capacitive coupling from gate to nanotube, which we find to be 31 pF/m.[29] Using an estimated coverage of 1.2 molecules/nm$^2$ for the case of DR1P,[5] and considering a typical threshold shift from Figure 3 of approximately 3 V, the net charge transferred by the molecules to the nanotube would have to be nearly 0.5 electrons/molecule to explain the typically observed shifts. Our DFT calculations, reported in Table 1, indicate that direct charge transfer (based on a Mulliken population analysis) from the chromophores to the nanotube is much too small to play an important role in determining $\Delta V_{th}$. For both *trans* and *cis* orientations of the chromophores, and for both parallel and perpendicular configurations, we find a maximum charge transfer per adsorbed molecule of 0.026 of an electron charge. It is interesting to note that the charge transfer in pyrene-functionalized SWNTs is comparable to the DFT results for the perpendicular configuration, where only the pyrene functional group is proximal to the SWNT. All of the small charge-transfer values are supported by Raman results shown in Figure 2e, where the undetectable G-band shift before and after functionalization confirms no significant charge transfer occurs between the chromophores and nanotubes.

There are different mechanisms that can lead to the observed threshold shifts, all of which appear to contribute to the observations presented in Figure 3. First, the observation of a threshold shift for pyrene-functionalized nanotubes, in addition to the chromophore functionalized devices, suggests that functionalization can displace gas molecules adsorbed on the nanotubes, of which oxygen is known to be one of the most important in causing p-type nanotube transport of the kind we see here.[36] An electrostatic gating effect, arising, e.g., from the the alteration of the metal work function through chemical modification, could also lead to corresponding threshold shift in the functionalized devices.[37] Other surface molecules have also been shown to play a role in determining nanotube FET hyteresis[38] and the observation of an average $\Delta V_{th}$ = -15 V for the backward sweep on DR1P devices, more than three times larger than that for DR1P', suggests that the type of SWNT functionalization also has an important impact on the hysteresis mechanisms. A preferred orientation for the chromophores – as previously observed in SHG measurements[23] – could lead to an effective gate field due to the orientation of the dipole moments,[5] which could be different for the two chromophores. In the absence of such a preferred orientation, the chromophore could impact the local dielectric environment and this screening may also depend on the chromophore dipole moment.

## 4. CONCLUSION

In summary, we have shown that a newly designed chromophore, DR1P', with opposite dipole moment to DR1P, was cleanly functionalized to SWNTs through the use of a pyrene tether, without any visible aggregation or clumping on the surface or on the nanotube. Raman spectroscopy confirms the presence of chromophores on the samples. Functionalized nanotube-based FETs were fabricated and characterized. Large on/off ratios for the SWNT transistors were preserved after functionalization with both DR1P and DR1P'. Intriguingly, the overall uniformity of the FET characteristics improves after functionalization. Threshold shifts for the transistor characteristics are observed and differ markedly (by more than a factor of 3) between chromophores with different dipole moments, providing opportunities for future device design and engineering.

AUTHOR INFORMATION

**Corresponding Authors**

* E-mail: pgopalan@wisc.edu (P.G.); maeriksson@wisc.edu (M.A.E.).

**Present Addresses**

[§]Departments of Chemistry and of Materials Science & Engineering, Drexel University, Philadelphia, Pennsylvania 19104, USA

**Notes**

The authors declare no competing financial interest.

ACKNOWLEDGMENT

We acknowledge financial support from the Division of Materials Sciences and Engineering, Office of Basic Energy Science, U.S. Department of Energy under Award #ER46590. B.M.W. acknowledges the National Science Foundation for supercomputing resources through the Extreme Science and Engineering Discovery Environment (XSEDE), Project No. TG-CHE130052.

ASSOCIATED CONTENT

**Supporting Information**

Additional details of DR1P' synthesis, NMR and UV/vis data, details on the AFM height profile analyses of the nanotube before and after functionalization, density of states (DOS) for SWNTs functionalized with DR1P and DR1P', and DFT calculations for the noncovalent interaction energy for chromophores to a (10,0) nanotube. This information is available free of charge via the Internet at http://pubs.acs.org

**Scheme 1.** Representation of the synthesis of DR1P' (compound 4).

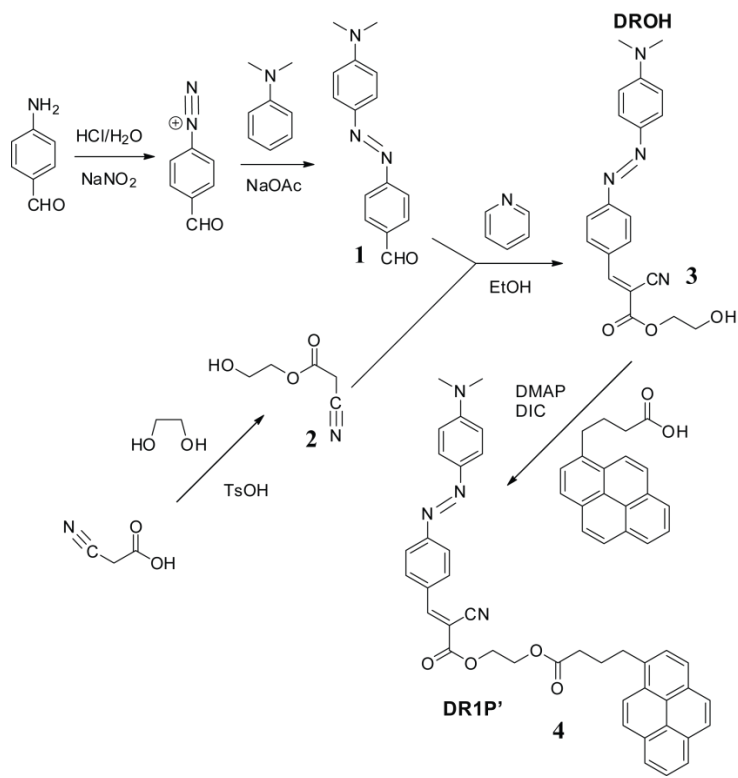

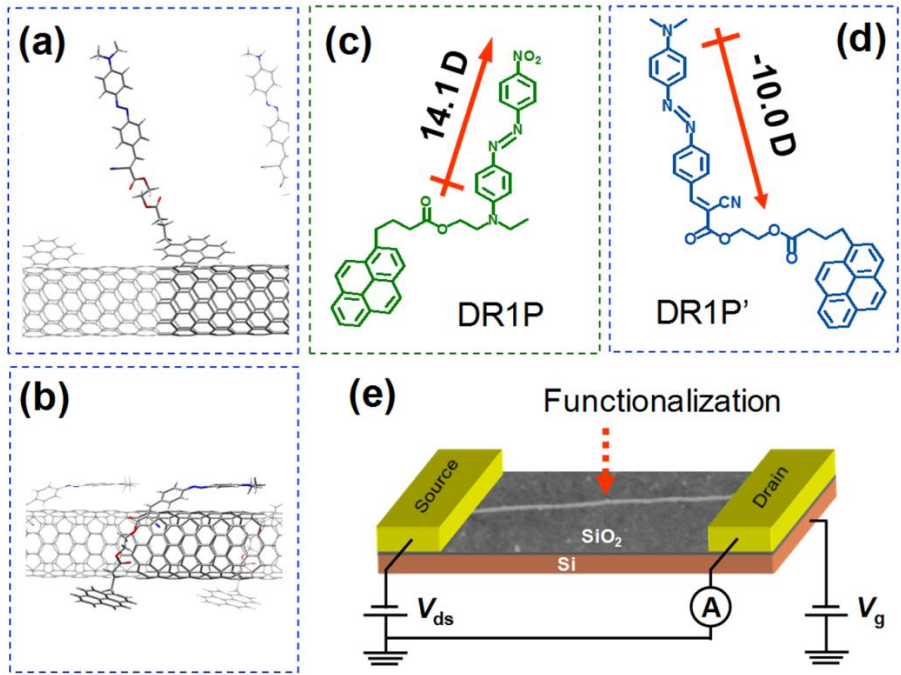

**Figure 1.** Optimized geometries of (a) perpendicular and (b) parallel geometrical orientations of DR1P' on a (10,0) nanotube from DFT calculations using the M06-L/6-31G(d,p) level of theory. Schematic molecular structures of the dipolar chromophores: (c) DR1P, with a dipole moment of 14.1 D; (d) DR1P', with an opposite dipole moment of -10.0 D. (e) SWNT field-effect transistor (FET); composite depiction for schematic purposes only: the schematic diagram here is formed from an AFM image combined with drawings of the source and drain, the 200 nm thick silicon dioxide layer, and the silicon substrate, which acts as a backgate. In the devices measured, the nanotube channel length $L_{ch}$ is 500 nm.

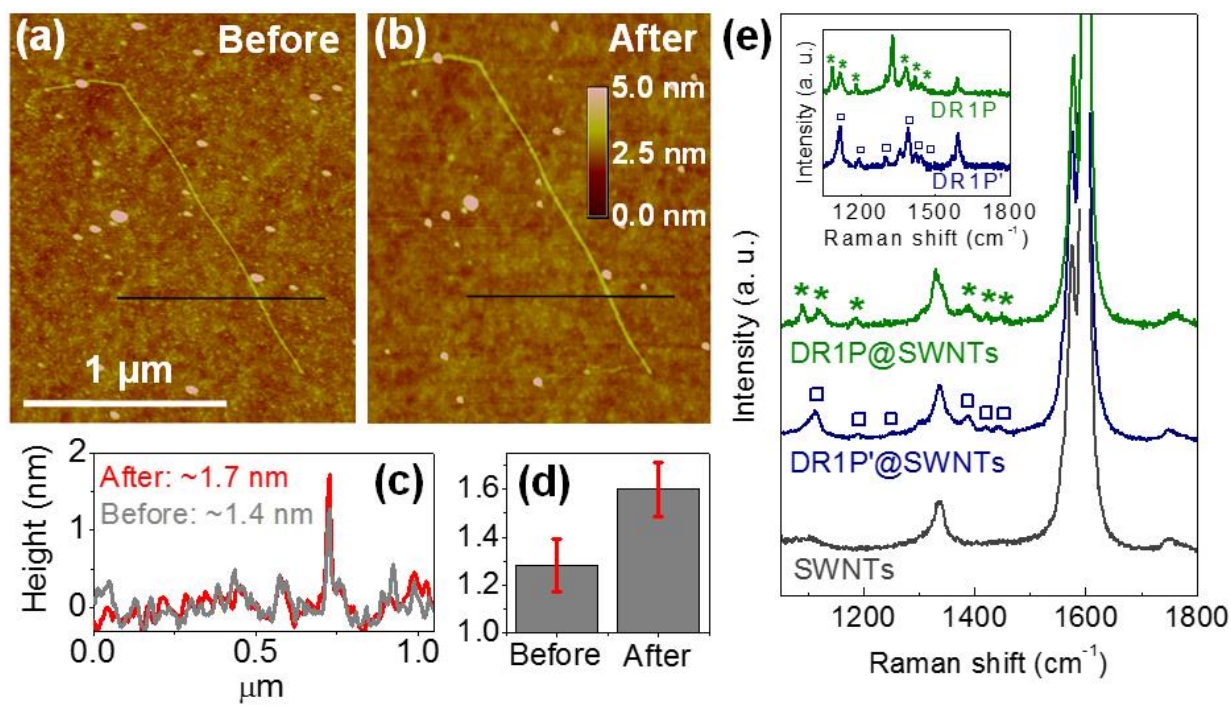

**Figure 2.** AFM images of an individual SWNT (a) before and (b) after DR1P functionalization. (c) Section analysis of the nanotube showing a height increase of ~ 0.3 nm before and after functionalization. (d) Statistical data based on section analysis of 8 different positions on the nanotube (see Supporting Information, Figure S4 and Table S1). (e) Raman spectra of pristine SWNTs, and the SWNTs functionalized with DR1P and DR1P', respectively. The inset shows the corresponding Raman spectra of DR1P and DR1P'.

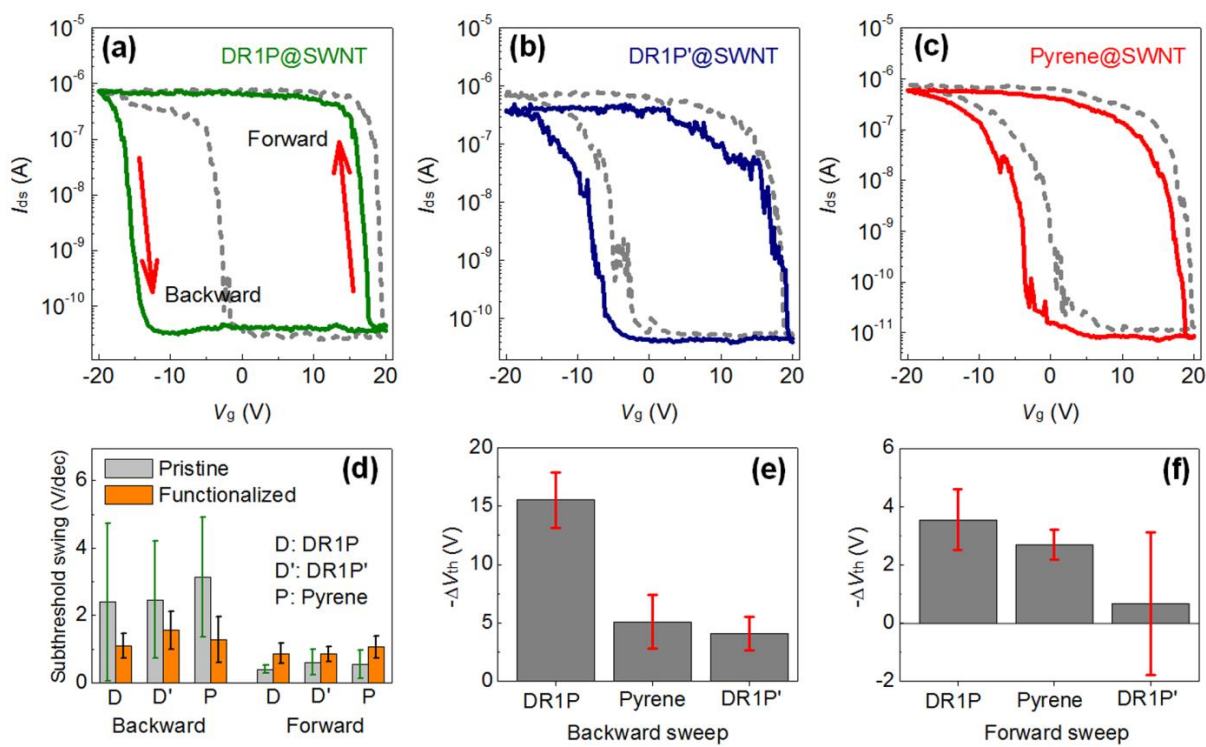

**Figure 3.** Transfer characteristics of the individual nanotube FETs before and after functionalization. Measurements were carried out in air with a source-drain bias ($V_{ds}$) of 100 mV. (a) Current-gate voltage ($I$-$V_g$) curves of a SWNT FET functionalized with DR1P, showing negative threshold shifts ($\Delta V_{th}$) in both forward ($V_g$: from 20 V to -20 V) and backward ($V_g$: from -20 V to 20 V) sweeps. The dotted line presents the $I$-$V_g$ curve of the corresponding pristine SWNT FET. Similar plots for the case of (b) DR1P' and (c) pyrene functionalization. (d) Statistics of subthreshold swings of a total of 23 nanotube FETs before and after functionalization, indicating a marked improvement in device uniformity induced by functionalization. (e) and (f) show measurements of $\Delta V_{th}$ in backward and forward $V_g$ sweeps of the SWNT devices functionalized with DR1P (7 devices), pyrene (8 devices) and DR1P' (8 devices), respectively.

**Table 1.** DFT calculations of the charge transfer to a (10,0) nanotube per molecule adsorbed in units of the absolute value of the electron charge. In all cases the calculated charge transfer per molecule is very small. Note that here we expect all molecules to be in the stable *trans* state; we report the results for the *cis* state for completeness.

| Orientation | DR1P | | DR1P' | |
| --- | --- | --- | --- | --- |
| | *trans* | *cis* | *trans* | *cis* |
| Parallel | 0.006 | 0.000 | 0.010 | 0.012 |
| Perpendicular | -0.017 | -0.018 | -0.026 | -0.026 |

SYNOPSIS TOC

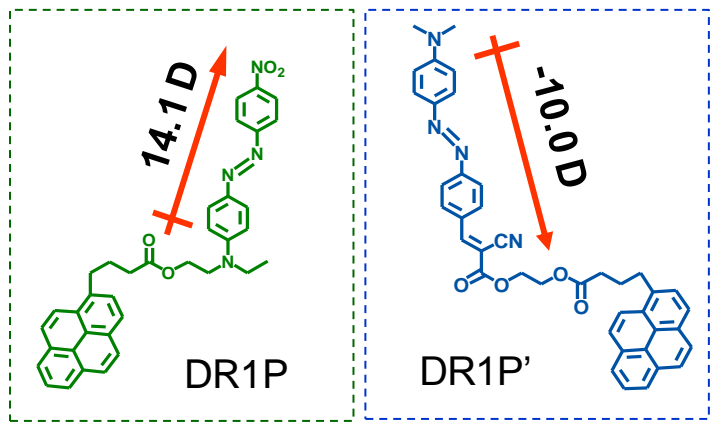

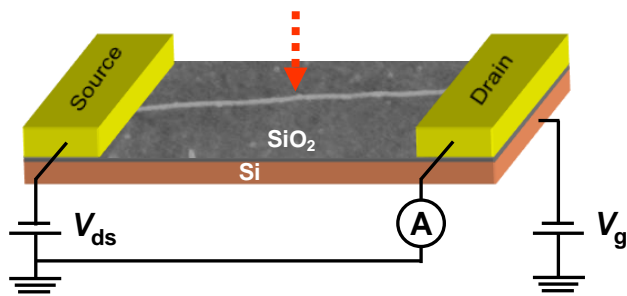